# A green program lifecycle supporting energy-efficient applications

Nadia Gamez[1], Jose-Miguel Horcas, Mónica Pinto[1], Lidia Fuentes[1]

[1] Dpto de Lenguajes y Ciencias de la Comunicación, Universidad de Málaga
{nadia, horcas, pinto, lff}@lcc.uma.es

## 1. INTRODUCTION

The carbon dioxide emissions dramatically increased during the past 10 years, accelerating the greenhouse effect. With the advent of the Internet of Things (IoT), the percentage of global emissions attributable to Information Systems is expected to further increase in the coming years, due to a proliferation of Internet-connected devices omnipresent in our daily lives (e.g., electric meters, wearable devices, etc.) [1]. Although software systems do not directly consume energy, they strongly affect the energy consumption of the hardware [2]. So, developers should be more aware of the energy consumed by these systems during their lifetime, and think about the long-term consequences in the sustainability of our planet Earth [3,4,5].

Indeed, once deployed, the energy consumed by a system depends on several factors determined mainly by the usage context. For example, it depends on the amount of data the system needs to store, transfer or query, and also on how the user interacts with the system. For example, initially, a stockbroker makes a more intensive use of the Stocks app installed in an iPhone comparing to an occasional investor user. But, the latter could intensify the use of this app after inheriting or selling a valuable property, modifying at runtime the initial behavior expected for that kind of user. So, the user behavioral pattern strongly impacts the final energy expenditure of applications. Therefore, **applications should be prepared at design time to be energy efficient during runtime** under real execution environments. This means that the area of energy-efficient software development needs green development lifecycles that provide appropriate methodologies and tools to identify and analyze the energy hotspots of applications early at design time, and see how they can be self-adapted to the runtime context usage.

Regrettably, there is a narrow view of developers and users and their responsibility in the energy consumed during application execution [3,4,5,6]. Users neither think about the environmental impact of their interactions with applications, nor do they know how their behavior can influence in saving energy. Developers rarely address energy efficiency as some recent studies show [3,4,5,6], mostly because they lack appropriate methodologies and tools that help them to produce green software at runtime. So, although software energy efficiency is becoming increasingly important in an ever more technology-dependent world, **development processes of self-greening systems supported by tools are still in their infancy** - see state of art sidebar-.

On the other hand, considering that many of current applications are normally deployed in smartphones or in any kind of smart objects (e.g., sensors, watches, etc.), optimizing the **energy consumption during the execution will also have a strong impact in battery saving**, enhancing the quality of experience of final users.

**Challenges**

- **The first challenge that should be addressed is to provide means to identify runtime energy hotspots**, i.e., which part of the code might has a great impact in the energy expenditure during system operation
- Since some energy hotspots could be bound at runtime to different components, so a great challenge is to **explicitly define the variability of design solutions that can mitigate the energy consumption according to current user interaction**
- **Provide developers with tools that help them to make a sensible eco-efficiency analysis at design time**, about the possibilities of optimizing energy consumption at runtime for a given application
- **Define energy reconfiguration rules to adapt the application to the varying usage patterns by exploiting the energy saving scenarios identified in the eco-efficiency analysis**
- **Provide a non-intrusive design and implementation solution that endows applications with self-greening capacities at low energy cost**

**Energy awareness in developing energy-efficient software (state of art)**

As the increasing number of papers that address energy consumption indicates, the software developer community is starting to pay more and more attention to the energy-efficiency concerns. Here we summarize some representative works of current research in this area.

**Empirical studies.** Recent empirical studies [3,4,5,6] made at different stages of the software life cycle, and with different groups of software developers, show that software developers do not have enough knowledge about how to reduce the energy consumption of their software solutions. Thus, the majority of developers are not aware about how much energy their application will consume and so, they rarely address energy efficiency [5,6]. Even practitioners that appear to have experience with green software engineering have significant misconceptions about how to reduce energy consumption [4]. These studies also evidence the lack of tool support of green computing, not only at the code level, but also at higher abstraction levels – i.e. requirements and software architecture levels [3]. The main conclusion of these studies is that software developers need more precise evidence about how to tackle the energy efficiency problem and some methodological and tool support that help them to effectively address it [4,6].

**Experimental works at code level.** There are plenty of experimental approaches that try to identify what parts of an application influence more in the total energy footprint of an application –i.e., to identify the energy hotspots [7]. These works propose to minimize energy consumption by focusing on code level optimizations. A common goal to all of them is the definition of energy profiles for different energy-consuming concerns. They usually focus on one particular energy-consuming concern and report the energy consumption of different implementations. For example, of data collections in Java [8], system calls in Android applications [9], http requests [10], or cloud applications [11]. Although some energy-consuming concerns have strong dependencies with others (e.g., to store data remotely will depend on the communication concern, and the latter one, on the security concern), the identification and management of these dependencies are still not appropriately covered by most of the existing works [12].

**Reasoning about energy efficiency at design level.** Although the experimental works at the code level are key to improve the knowledge of software developers regarding the energy consumption of software systems, there are other works that demonstrate that changes at the design level tend to have a larger impact in energy consumption [13,14]. These works consider energy consumption as a new quality attribute [15]. For instance, sustainability is proposed as a quality attribute in [15], focusing specifically on the resource consumption sub-characteristic that is decomposed into the software utilization, energy usage and workload energy quality properties. What is important at this level is to be able to compare the energy consumed by different design alternatives, and also to be able to perform a tradeoff between energy efficiency and other quality attributes [15]. There are some relevant approaches that focus on the design of catalogs of energy-aware design patterns [16] and the definition of architectural tactics [15], as well as new architecture description languages that incorporate an energy profile and analysis support [17,18]. The experimental part of these works consists of checking at the code level the effects of applying specific design or architectural patterns [15,16].

**Energy-based reconfiguration at runtime level.** Here we focus on proposals that are able to monitor changes on the user behavioral patterns and react to the effects of those changes on the consumption of energy. They should also be able to update the behavior of applications to their 'energy usage profile'. The final goal is to maintain the energy consumption of the software system within reasonable levels. Some proposals monitor the energy consumption of previously identified energy hotspots at runtime [7], and others build real-time profiles of energy consumption [19]. A few of them apply a models@runtime approach to introduce the concept of energy consumption and use the energy information to adapt the software during runtime [20]. There are examples of the dynamic reconfiguration of energy aware software in different domains. For instance, in [21] the energy efficiency of mobile applications is improved by analyzing and deciding at runtime if a 'hot' part of the application has to be offloading to be executed on a cloud-based server. Also, in [22] an adaptation model is proposed to minimize energy consumption of cloud applications at runtime. Finally, the work in [23] presents DREAMS, a Dynamically Reconfigurable Energy Aware Modular Software architecture for sensor networks. None of them defines a generic and reusable approach as we pretend to make.

## 2. HADAS GREEN LIFECYCLE OVERVIEW

The development of self-greening applications can have a significant impact on decreasing energy consumption. Therefore, developers need methods and tools to generate eco-efficient applications (i.e., applications that make an optimal use of resources), which later can be dynamically adapted to optimize the real energy consumption considering the user interaction at runtime. The key to achieve this is to model the energy as a quality attribute, giving it priority over other traditional quality attributes such as performance or response time. Therefore, we argue that the responsible of driving a green system development, during all the software life cycle, including runtime, should be the energy attribute.

There are still some open challenges -see the sidebar- we need to afford in order to develop real energy-efficient applications. In this work, we present HADAS, a development process of self-greening applications. The green program lifecycle of HADAS proposes to collect energy-related information at design time and use it at runtime to adapt the application behavior to the real energy consumption.

Figure 1 shows the main steps of the green lifecycle we are proposing. Firstly (Figure 1, label 1), at design time the developer has to identify, in the application requirements, the energy hotspots that can be adapted at runtime. In our approach, a *runtime energy hotspot* is a point in the application that under certain conditions can consume much energy and if these conditions change at runtime it is possible to reduce this energy consumption by modifying the application running components. Some examples of these hotspots are Store (store and retrieve data) and Communication (send short or large messages with more or less frequency). The concerns that model the runtime energy hotspots at design time can be considered as *energy consuming concerns*, which could be designed in different ways. For example, there are different options to store data (in a data structure, cache memory, etc.), each with a different energy consumption that depends on some input parameters such as the size or type of data. All the alternative design solutions for every energy consuming concern are previously stored in the HADAS Green Repository, which purpose is to encapsulate the knowledge about what implementation solutions consume more than others. Then, the application developers will select the energy consuming concerns identified as part of the application functionality (e.g., store, compression) and the variants they want to explore (e.g., variants for store can be to store data in a local file or in a server). This selection is done using the HADAS Assistant Tool, which offers several Google Forms. As result, HADAS gives the expected energy consumption of all the options selected for every energy consuming concern in several graphics (Figure 1, label 2). This useful information is then used by the application developer to make a **Sustainability Analysis** at design time and choose the preferred options. HADAS then helps the developer to take informed decisions about the energy consumption of the selected concerns, and generate the initial application configuration. This sustainability analysis will also help to identify those situations where the energy expenditure strongly depends on some parameters that can vary at runtime. This information will be used by the developer to specify the self-greening rules that will trigger a reconfiguration at runtime.

Later, once the application is running, HADAS proposes a solution to perform a runtime **Energy Analysis** that consists of detecting any relevant variation in the parameters that have influence in the power consumption of the energy consuming concern, as the file sizes (Figure 1, label 3). This is done by an energy-efficient context observation that should be implemented independently of the application base architecture and with a low energy cost. Then, when a context change impacting the energy consumption is detected, HADAS will carry out an **Energy Optimization** by reconfiguring the application. This consists of modifying the variants of the selected energy consuming concerns by other variants more adequate to the new usage of the application. After this, by modifying the modules implementing the energy consuming concerns, the new application configuration is now more energy efficient, adapted to the current application parameters.

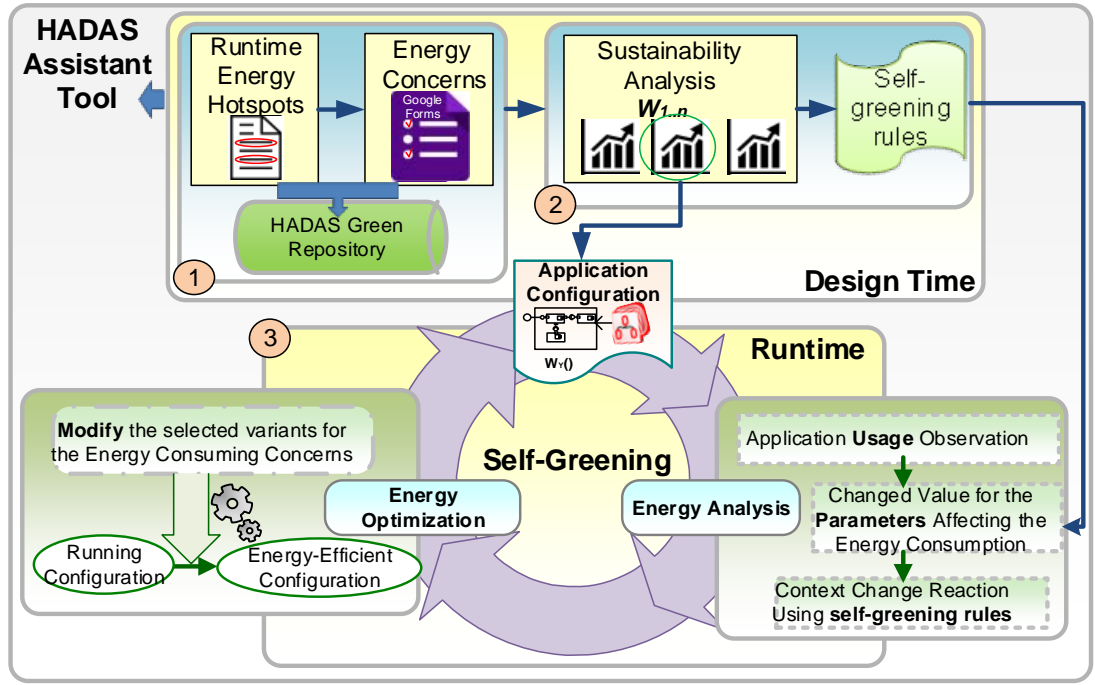

**Figure 1.** HADAS Green Lifecycle

In the rest of the paper we will provide more details about each step of our proposal using a comprehensible example, a Media Store (MS) system for mobile applications.

## 3. MODELING RUNTIME ENERGY CONSUMING CONCERNS

Predicting and simulating the energy expenditure of a given application can give hints about the final power consumption of the application. However, the energy consumption strongly depends on several factors, and some of them will vary at runtime. For example, some users of instant message applications make a more intensive use of audio messages and others prefer to write textual ones. The message type, data format or how frequent the user uses some components can give opportunities to optimize the energy expenditure of our device. But, before identifying how we can optimize the energy consumption at runtime, we should focus on identifying the runtime energy hotspots.

So, the green lifecycle we propose starts at design time by identifying which are the energy consuming concerns that most affect the runtime energy expenditure. After analyzing several works we can conclude that many energy consuming concerns are recurrent, appearing in the majority of applications. Usually they are scattered or crosscut several components (i.e., they are crosscutting concerns) [24], so it would be beneficial to model and implement them, independently of the system functionality, to facilitate their replacement at runtime by more eco-efficient designs or implementations. Since these concerns are usually present in many applications we propose to store them in the HADAS Green Repository ready to be reused by any application.

The developer will identify first those parts of the application that will impact more the final power consumption. But, instead of making this from scratch, the developer can do it by performing a mapping between the energy consuming concerns modeled as part of the HADAS repository and the concerns identified as runtime energy hotspots the application has to implement. Likewise designers are able to identify which part of the application demand a particular design pattern, now designers have to develop a sense of smell to identify the concrete runtime energy hotspots of their applications. Learning to recognize energy hotspots is absolutely necessary and helpful in any energy-aware development process.

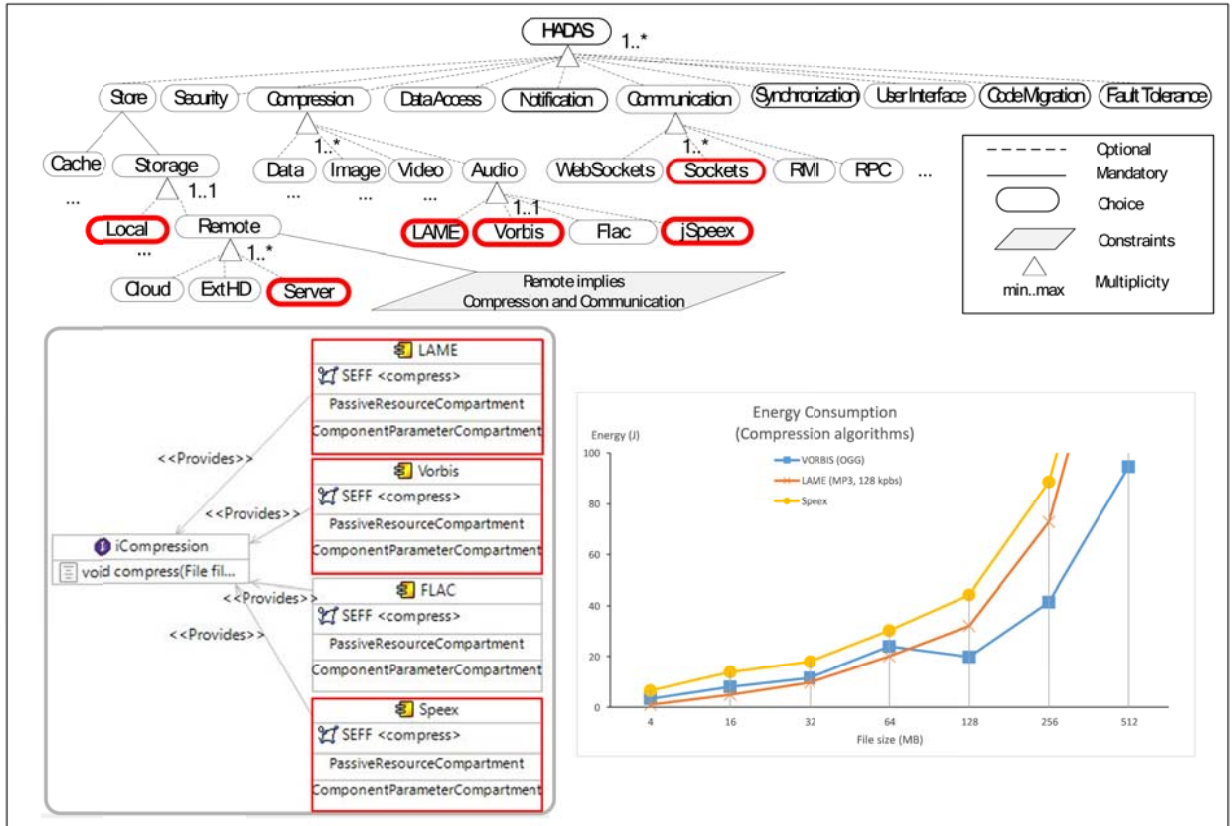

**Figure 2.** Models of the HADAS Repository and Power Consumption graphic of audio codecs

There are plenty of studies that show there is a high variability of alternative implementations and design solutions to many energy consuming concerns [8,9,10,11,12], and some of them permit their replacement at runtime to achieve energy savings. Then, HADAS Green Repository contains a set of models that encapsulate different solutions for each of the energy consuming concerns. We explicitly model the variability of energy consuming concerns using a variability model, concretely CVL [25]. The motivation to use CVL is that it allows to easily maintaining connections between energy consuming concern variants specified in a Vspec tree, and the set of components that implement this variant. The top of Figure 2 shows an excerpt of the VSpec tree with some energy consuming concerns like Store, Communication, Compression, Security, Data Access, etc. Focusing on data compression, which is one of the concerns present in the MS application, we include several algorithms that consume more or less energy (e.g., LAME, Vorbix and jSpeex audio codecs) depending on file size.

So, what the developer needs to know at design time are the options that exist to address a concrete runtime energy consuming concern, and the expected energy consumption of each of them at runtime. Energy consumption mainly depends on the resources that each application component is expected to consume (e.g., cpu cycles, disk access, etc.) for different variable parameters and on the hardware characteristics (e.g. cpu cycles/s, MB/s, etc.). With this information it is possible to calculate the expected energy consumption by conducting experimental studies, or by estimation using some energy models and simulation. For HADAS, the concrete amount of joules consumed by different energy consuming concerns considering concrete hardware is not so important, but the relative energy it is. So, the intention of HADAS is to store the energy consumption obtained following different approaches, and provide this information to the developer. Then, we can collect the results of many of the experimental studies that are being published, store them in the HADAS repository and provide advice based on these results, along with an initial design for each of the possible solutions. The energy consumption showed in this paper was experimentally calculated, but we have also explored the use of the Palladio toolset [26] and its Power Consumption Analyzer (PCA) [18], an IDE perfectly well suited for predicting by simulation the energy consumed by an architecture design. Indeed, the component model showed in Figure 2 is based on PCM (Palladio Component Model).

The main benefit of Palladio and other similar frameworks is that it is possible to simulate energy consumption for different hardware. However the developer has to manually estimate the resources to simulate the expected energy consumption for each operation of the components that are part of a given solution. Anyway, whatever the approach used to calculate the expected energy consumption, the effort of measuring, estimating and/or simulating the energy expenditure of each of the possible energy consuming concerns would be an intractable task for developers. Note that for instance ten concerns the number of alternative solutions would be 100 on average. So, the goal of HADAS is to save time to application developers by doing this manual and tedious job and putting the results in the HADAS repository. Of course this is a collaborative task thought to be performed incrementally.

Coming back to our example, the energy consumption for each audio codec variant was experimentally calculated for the MS system. Figure 2 bottom right, shows the graphic with the power consumption to compress 9 WAV audio files of different sizes (from 4Mb to 512 MB) using the following audio compression algorithms implemented in Java: Java LAME 3.99.3 to create MP3 audio files using a bitrate of 128Mb, Vorbis-java (libvorbis-1.1.2) to compress in OGG files, and Java Speex Encoder v0.9.7 to compress in SPX files. The energy expenditure is measured with JouleMeter [27], a Microsoft tool to measure the energy usage of software applications running on a computer. We repeated each experiment 5 times and use the median in Joules that appears in the graphic. This tool has been calibrated using Watts'Up [28] to obtain the real power consumption depending on every hardware component (e.g., CPU, HDD, Screen, ..). All the experiments have been conducted in a Desktop PC with Intel Core i7 CPU, 3.4GHz, 16 GB of RAM under Windows 10, 64 bits.

Now imagine we want to calculate the resource consumption of sending a compressed file through a TCP socket, which entails the compression and communication concerns. HADAS formally specify the dependency relationships between energy consuming concerns using the cross-tree constraints of the variability model. As previously explained, the variability model includes all the variants for all the energy consuming concerns of the repository. For example, the Remote Storage concern depends on both the Communication and the Compression concerns, so we define a cross-tree constraint associated to the Remote feature as: *Remote implies Compression and Communication*. With HADAS, designers do not need to be aware about the inter-dependencies between the concrete solutions of different energy consuming concerns. HADAS will enable and disable variants of different hotspots automatically, as the designer selects the desired options. For example, if developers select to store files in a server, they do not need to be aware of what other energy consuming concerns need to be included in the sustainability analysis. In the example, HADAS automatically incorporates the Communication and the Compression concerns because they are also energy consuming.

So, the developer of the MS application has to reason about different audio codecs to store files locally or in a server (see the corresponding features highlighted in red). In next section we will show how HADAS graphics are used to perform a sustainability analysis at design time that helps developers to define the runtime reconfiguration rules.

## 4. ANALYSING AND SELECTING ENERGY-EFFICIENT CONFIGURATIONS

The key to success of self-greening applications is to fully exploit the energy saving possibilities arisen at runtime. So, the main role of HADAS Green Repository in the development of self-greening applications is to provide the necessary means to make an energy-efficiency analysis at design time, about the possibilities of optimizing energy consumption at runtime for a given application. This means that the HADAS Green Repository can be used to see if it is worthy to specify a reconfiguration rule that replaces at runtime a concrete concern implementation by other one after, for instance, a drastic user behavior modification. So, the HADAS toolkit helps developers to make a comparative analysis of the power consumption of different solutions for a given runtime energy hotspot. For example, in Figure 2 we can see that for a file size equal to 4MB all the codecs consume similar energy, so can deploy the LAME codec, but when this size increments up to 64MB, then the codec Vorbis is greener. Since both the file size and quality depend on what the user needs in every moment, it is not enough to generate an initial energy-efficient application, but to codify reconfiguration rules to replace one solution when the current one is no longer the most energy-efficient, under the current use conditions (e.g., audio codec LAME by Vorbis).

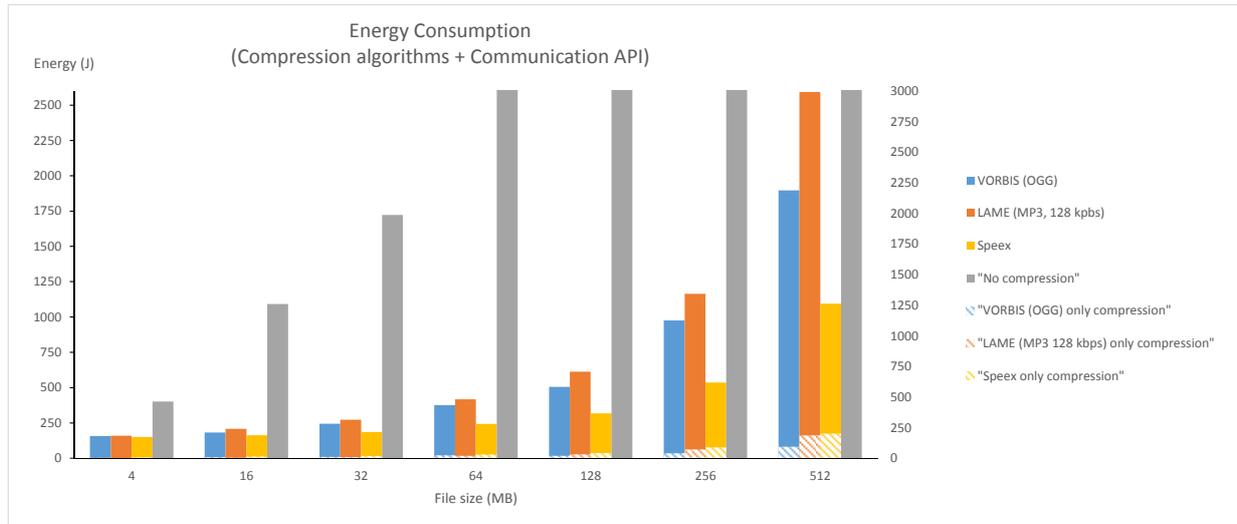

**Figure 3.** Energy consumption graphic of compression and communication concerns

With HADAS the developer is aware of that the decision of choosing an audio codec can only be made considering the expected use of the application. This reasoning may be described using Event Condition Action (ECA) rules [29], a simple but efficient reconfiguration mechanism that consumes less than other computationally more complex approaches like, for example, optimization algorithms. In our case the event will be a variation in the parameter value that affects the energy expenditure of a given concern (e.g., file size); the condition will be the concrete value that makes the current energy consuming concern implementation no longer optimal (e.g., size > 64Mb); and, the action will be to replace current component configuration by a more eco-efficient solution (e.g., replace LAME by Vorbis).

But, this reasoning cannot be performed in isolation for every energy-consuming concern, because reducing the energy of one concern can have a collateral effect of incrementing the energy expenditure of others. In the MS application, we have already said that the developer is also interested in exploring the possibility of uploading the audio files to a server. In this scenario, audio files must be first compressed and then uploaded to a server. In this case we need to know the total power consumption of compressing the file and sending it to the server. Notice that different compression algorithms produce compressed files of different sizes, and therefore the energy consumed by the communication concern will be different depending on the compression algorithm previously used. HADAS will help developers to jointly reason over different concerns, by showing the graphics with the power consumption for the entire configuration. The configuration is generated according to the dependency relationships previously defined in CVL (Figure 2). For our example, Figure 3 shows the power consumption considering the two concerns used in the remote-server configuration: the Compression and the Communication concerns. This graphic shows that for a file size of 4 MB the energy consumption of the three audio codecs plus communication is similar, but as the file size increases, the greener codec is Speex.

With all this information the developer can now complete the reconfiguration rules for the MS application. Since the great majority of MS users will store typical song audio files of 4 MB, then the developer can select the local feature (i.e., store audio files in the device) and the LAME codec (i.e., the greenest according to Figure 2) for the initial configuration. However, some users may be interested in storing audio files with a size greater than 64 MB (e.g., a journalist that wants to record an interview or somebody that wants to record some performance). In this latter case the greenest codec is the Vorbis (i.e., the greenest according to Figure 2). Finally, as the file size increases the device memory becomes full, so it's time to upload the audio files to a server.

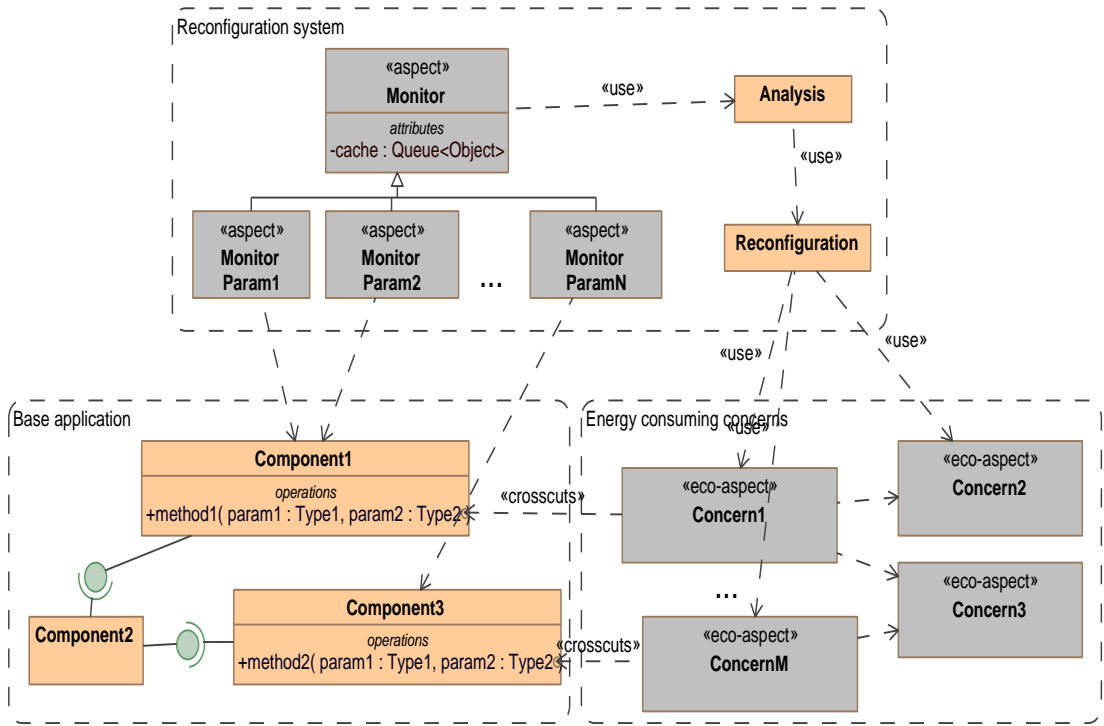

**Figure 4.** HADAS self-greening architecture

But, according to the results showed by HADAS (Figure 3), the energy consumption of sending the file to the server increases exponentially in function of file size, and thus, a greener solution is to change the compression algorithm by other with a bigger compression rate, the Speex codec. Note that the Speex codec is the compression algorithm that consumes the most if it is used locally (see Figure 2). There are different green solutions because the file size affects communication to a greater extent than it does in compression as shown in Figure 3. So, reducing the file size as much as possible before sending it to the server drastically decreases the energy consumption of the global solution. This means that we need an additional reconfiguration rule that specifies that if the user or the system decides to upload the audio files to the cloud, the greenest codec is Speex. So, we have identified three energy saving scenarios at runtime, each one recommending a different audio codec. In next section, we will show a possible implementation of a self-greening application written in Java.

## 6. ENERGY-AWARE RECONFIGURATION

Once the initial system configuration is deployed in distributed devices, the system has to monitor and reconfigure the current system, pursuing truly energy efficiency at runtime. But, how can we implement a self-greening application without overloading the system with heavy-energy monitoring mechanisms? What elements should be monitored at runtime? How can we analyze the context to enforce a self-greening behavior without complicating the resulting code?

Indeed, the great challenge is to define a self-greening mechanism that wastes the least amount of energy, being then not recommended to apply burdensome self-adaptation approaches (e.g., manipulating models@runtime [30]). In addition, since eco-efficient concerns crosscut several application components it makes sense to implement energy-related concerns separately of the application functional components, to facilitate its replacement at runtime.

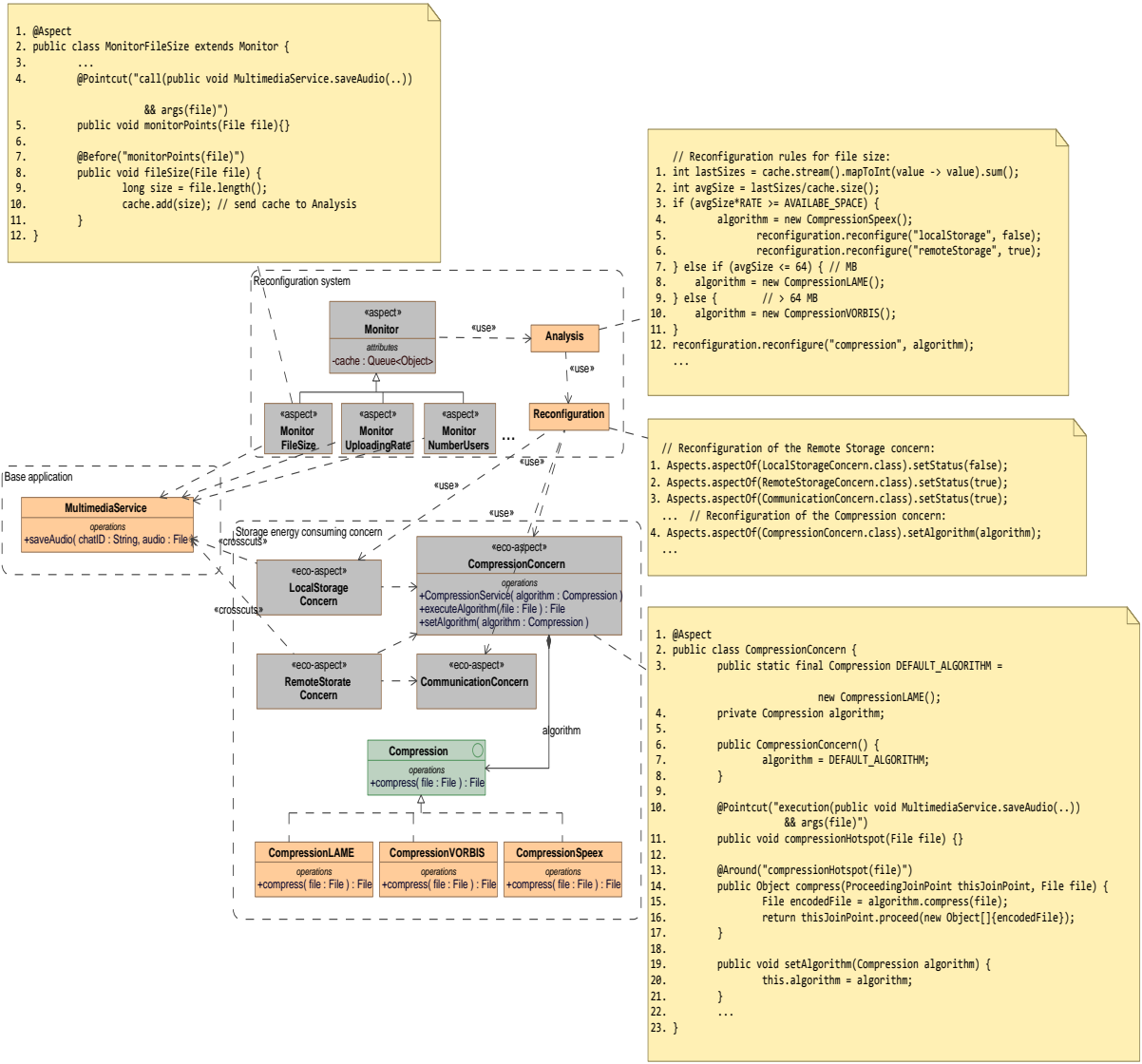

**Figure 5.** Implementation of the media store using HADAS self-greening approach

There is no single possible implementation of the classical self-adaptation MAPE-K loop [31] (monitoring, analyse, plan, execution and knowledge) in Java, but not all of them fulfill the two requirements of consuming low energy and being not intrusive. Since we need to observe the runtime variation of some parameters, the subject-observer design pattern could be a good option, and so is the use of Java events. However, we have found that a solution that is both, not intrusive and eco-efficient, is the AspectJ language, an Aspect-Oriented (AO) extension of Java. With this language it is possible to define interception points in the application base code where we want to inject an extra-functional property, like the energy consuming concerns in our case [24]. Before, around or after executing these interception points (i.e., pointcuts in AspectJ terminology) we can inject code related to self-greening functionality separately from the core application code. The AspectJ annotations are interpreted at compile time by the ajc compiler that weaves the "aspect" code with the application classes at the bytecode level, so there is no overhead at runtime.

Figure 4 shows a generic aspect-oriented design solution for implementing self-greening applications in AspectJ that uses Java annotations to codify the self-greening functionality. A possible solution could be to implement the monitoring of events that will trigger a reconfiguration as separated code that will be injected in the base code of the application. At runtime we only need to observe those parameters whose variation implies that the current configuration is no longer the most energy efficient; i.e., these parameters are the events that appear in the ECA rules defined above (the file size in our case). So, we propose to implement a Monitor for each of the parameters to be observed as an aspect, i.e., annotated with @Aspect (Figure 4, stereotyped as <<aspect>>).

The monitored value captured by each monitoring class is sent to the *Analysis* component that contains the ECA rules to decide whether or not a reconfiguration is needed. If the rules determine that a new configuration is greener, the *Analysis* component will send the new configuration to the *Reconfiguration* component. This component directly interacts with the energy consuming concerns by enabling/disabling them and reconfiguring their internal behavior (e.g., replace an algorithm). The runtime energy consuming concerns are implemented also as aspects (i.e., ConcernI with stereotype <<eco-aspect>>) and are injected (i.e., see <<crosscuts>> link) into the base application code in a non-intrusive way. This provides a light solution in terms of energy consumption and allows an easier reconfiguration of the energy consuming concerns, because aspects do not call or are called by the base application.

Coming back to our example, Figure 5 shows how we injected the self-greening code into the base application using AspectJ. One of the monitor classes we have defined is the *MonitorFileSize* that monitors the size of the audio files processed by the *MultimediaService.saveAudio()* method (line 4 in the *MonitorFileSize* aspect). Every new parameter value is stored in a cache (line 10), so as the *Analysis* component can have information about the most recently activity of the user and take more accurate decisions. In our example, the Analysis considers the average size of the latest files processed (line 2 in the *Analysis* component) in the reconfiguration rules associated with the file size. The rest of the code of this component shows the implementation of the ECA rules defined in the previous step. Lines 7-8 correspond to the ECA rule for files stored locally and with a size less or equal to 64 MB, setting the LAME codec. Lines 9-10 implements the second ECA rule that sets the Vorbis algorithm for file sizes greater than 64 MB. But, when the space available in the local repository is almost full, the system will be reconfigured to store the files in a remote server and thus, change the compression algorithm by Speex (lines 3 to 6).

The *Reconfiguration* component will activate and/or deactivate the appropriate concerns changing from the local to the remote storage configuration (lines 1 to 3 in *Reconfiguration* component). Also, it is in charge of changing the current configuration of the activated concern such as changing the compression algorithm (line 4). The energy consumption concerns crosscut the base application to inject the appropriate functionality at the correct place. For instance, the *CompressionConcern* aspect crosscuts the base application to compress the audio before saving it (lines 10 to 17 in the *CompressionConcern* aspect).

We tested our implementation and the AspectJ mechanism and the results showed that the energy consumption of the proposed implementation is insignificant comparing to the total amount.

## 7. CONCLUSIONS

We have presented HADAS, a self-greening approach that aims to optimize the energy consumption of applications at runtime. We focus on those concerns whose consumption depends on parameters that could vary at runtime in function of the user interaction or on other context information (e.g., available memory, battery level). In order to specify the self-greening rules we have develop a runtime energy consuming concerns repository with information about relative energy consumption of some recurrent concerns. The graphics generated by the HADAS Green Repository are used to analyse the possibilities of optimizing energy consumption at runtime. Indeed, we have shown that there are valuable opportunities to optimize the energy consumption at runtime that should not be neglected by developers. In our example, if the initial codec LAME is maintained and the user starts producing files greater than 64 MB we miss the opportunity to save between 48% (128 MB) and 65% (512 MB). Also, if audio files have to be uploaded to a server in a certain moment, setting the codec to Speex could save between 52% (difference with LAME) or 43% (difference with Vorbis) for files greater than 64 MB and as the file size increases the saving can be up to 81% (difference with LAME) or 54% (difference with Vorbis). To get this profit it is necessary to implement lightweight self-adaptation mechanisms with low energy consumption as the one used here. As part of our future work, we plan to extend the HADAS green repository with new concerns incorporating the new experimental results accomplished by us or by others. Also we plan to implement a new aspect weaver able to interpret reconfiguration rules defined at runtime.